\documentclass{PoS}

\usepackage{amstext}
\usepackage{amsfonts}
\usepackage{amssymb}
\usepackage{color}
\usepackage{caption}         
\usepackage{cite}
\usepackage{dsfont}
\usepackage{epsfig}
\usepackage{mathtools}
\usepackage{multirow}
\usepackage{physics}
\usepackage{ulem}
\usepackage{subcaption}

\title{Computation of hybrid static potentials from optimized trial states in SU(3) lattice gauge theory}

\ShortTitle{Computation of hybrid static potentials from optimized trial states}

\author{\speaker{Christian Reisinger}, Stefano Capitani, Lasse M\"uller, Owe Philipsen, Marc Wagner\\
        Goethe-Universit\"at Frankfurt \\
        Max-von-Laue-Stra{\ss}e 1\\
        Institut f\"ur Theoretische Physik\\
        D-60438 Frankfurt am Main, Germany\\
        E-mail: \email{reisinger@itp.uni-frankfurt.de}, 
        \email{capitani@itp.uni-frankfurt.de}, 
        \email{lmueller@itp.uni-frankfurt.de}, 
        \email{philipsen@itp.uni-frankfurt.de}, 
        \email{mwagner@itp.uni-frankfurt.de}}

\abstract{We compute hybrid static potentials in SU(3) lattice gauge theory using a method to automatically generate a large set of suitable creation operators from elementary building blocks. This method allows to find sets of creation operators, which generate trial states with large ground state overlaps for all investigated angular momentum and parity sectors. We present numerical results for hybrid static potentials with quantum numbers $\Sigma^-_g, \Sigma^+_u, \Sigma^-_u, \Pi_g, \Pi_u, \Delta_g, \Delta_u$.}

\FullConference{The 36th Annual International Symposium on Lattice Field Theory - LATTICE2018\\
		22-28 July, 2018\\
		Michigan State University, East Lansing, Michigan, USA.}

\begin{document}


\section{Introduction}

Mesons, where the gluons contribute to the quantum numbers in a non-trivial way, are called hybrid mesons. Such hybrid mesons are not limited to quark model quantum numbers $J^{\mathcal{P} \mathcal{C}}$ with $\mathcal{P} = (-1)^{L+1}$ and $\mathcal{C} = (-1)^{L+S}$, where $L \in \{ 0,1,2,\ldots \}$ is the orbital angular momentum and $S \in \{ 0,1 \}$ is the quark spin. Improving our understanding of hybrid mesons and other exotic hadrons is an important goal in modern theoretical particle physics (for a review cf.\ \cite{Meyer:2015eta}). For example, since hybrid mesons are also a popular topic for current and future experiments (e.g.\ the GlueX and the PANDA experiments; for a review cf.\ \cite{Olsen:2017bmm}), a sound theoretical knowlegdge is mandatory to interpret the expected experimental data.

In this work we discuss, how to compute hybrid static potentials with SU(3) lattice gauge theory and show first numerical results (for previous similar work cf.\ e.g.\ \cite{Juge:1997nc,Michael:1998tr,Bali:2000vr,
Juge:2002br,Michael:2003ai,Bali:2003jq,
Wolf:2014tta,Reisinger:2017btr}). Such hybrid static potentials are relevant for hybrid mesons with a heavy quark $Q$ and a heavy antiquark $\bar{Q}$, typically $Q \bar{Q} \in \{ c \bar{c} , b \bar{b} \}$. Quite often lattice gauge theory results for hybrid static potentials are used as input in effective field theory approaches or phenomenological studies, e.g.\ to predict the spectrum of heavy hybrid mesons in the Born-Oppenheimer approximation (cf.\ e.g.\ \cite{Braaten:2014qka,Berwein:2015vca,Oncala:2017hop,
Brambilla:2017uyf,Brambilla:2018pyn}). The focus of this work is on finding suitable sets of creation operators, which generate trial states with large ground state overlaps.


\section{\label{sec:1}Quantum numbers and trial states}

We compute hybrid static potentials from Wilson loop-like correlation functions using SU(3) lattice gauge theory. The gluonic excitations are realized by replacing the straight spatial Wilson lines of the Wilson loops by parallel transporters, which have a more complicated structure. We put the static quark and the static antiquark, which we treat as spinless color charges, on the $z$ axis at positions $\mathbf{r}_Q = (0,0,+r/2) \equiv +r/2$ and $\mathbf{r}_{\bar{Q}} = (0,0,-r/2) \equiv -r/2$.

Hybrid static potentials are characterized by the following quantum numbers:
\begin{itemize}
\item $\Lambda \in \{ \Sigma \doteq 0, \Pi \doteq 1, \Delta \doteq 2,\ldots \}$, the absolute value of the total angular momentum with respect to the $Q\bar{Q}$ separation axis, i.e.\ the $z$ axis.

\item $\eta \in \{g \doteq +,u \doteq -\}$, the eigenvalue of the combination of parity and charge conjugation $\mathcal{P} \circ \mathcal{C}$.

\item $\epsilon \in \{+,-\}$, the eigenvalue of the spatial reflection along the $x$ axis, which is an axis perpendicular to the $Q\bar{Q}$ separation axis, $\mathcal{P}_x$.
\end{itemize}
Note that for angular momentum $\Lambda > 0$ the spectrum is degenerate with respect to $\epsilon = +$ and $\epsilon = -$. The labeling of states is thus $\Lambda^{\epsilon}_{\eta}$ for $\Lambda = 0 = \Sigma$ and $\Lambda_{\eta}$ for $\Lambda > 0$.

A hybrid static potential trial state with quantum numbers $\Lambda^{\epsilon}_{\eta}$ can be constructed via
\begin{eqnarray}
\label{eq:trialstate} & & \hspace{-0.7cm} \ket{\Psi_\text{hybrid}}_{S;\Lambda_\eta^\epsilon} \ \ = \ \ \bar{Q}(-r/2) a_{S;\Lambda_\eta^\epsilon}(-r/2,+r/2) Q(+r/2) \ket{\Omega} ,
\end{eqnarray}
where
\begin{eqnarray}
\nonumber & & \hspace{-0.7cm} a_{S;\Lambda_\eta^\epsilon}(-r/2,+r/2) \ \ = \\
\nonumber & & = \ \ \frac{1}{4} \sum_{k=0}^3 \textrm{exp}\bigg(\frac{i \pi \Lambda k}{2}\bigg) R\bigg(\frac{\pi k}{2}\bigg) \bigg(S(-r/2,+r/2) + \eta S_{\mathcal{P} \circ \mathcal{C}}(-r/2,+r/2) + \\
\label{EQN644} & & \hspace{0.675cm} \epsilon S_{\mathcal{P}_x}(-r/2,+r/2) + \eta \epsilon S_{(\mathcal{P} \circ \mathcal{C}) \mathcal{P}_x}(-r/2,+r/2)\bigg) .
\end{eqnarray}
$R$ denotes a rotation around the $Q \bar{Q}$ separation axis, $S$ is a path of spatial links different from a straight line connecting the quark and the antiquark and $S_X$ denotes $X \in \{ \mathcal{P} \circ \mathcal{C} , \mathcal{P}_x , (\mathcal{P} \circ \mathcal{C}) \mathcal{P}_x \}$ applied to $S$.


\section{\label{SEC567}Correlation functions and lattice setup}

We determine hybrid static potentials with quantum numbers $\Lambda_\eta^\epsilon$, which we denote by $V_{\Lambda_\eta^\epsilon}(r)$, from the asymptotic exponential behavior of temporal correlation functions
\begin{eqnarray}
\nonumber & & \hspace{-0.7cm} W_{S,S';\Lambda_\eta^\epsilon}(r,t) \ \ = \ \ \bra{\Psi_\text{hybrid}(t)}_{S;\Lambda_\eta^\epsilon} \ket{\Psi_\text{hybrid}(0)}_{S';\Lambda_\eta^\epsilon} \ \ = \\
\nonumber & & = \ \ \bigg\langle \textrm{Tr}\Big(
  a_{S';\Lambda_\eta^\epsilon}(-r/2,+r/2;0)
  U(+r/2;0,t)
  \Big(a_{S;\Lambda_\eta^\epsilon}(-r/2,+r/2;t)\Big)^\dagger \\
\nonumber & & \hspace{0.675cm} U(-r/2;t,0)
\Big) \bigg\rangle_U \ \ \sim_{t \rightarrow \infty} \\
\label{EQN599} & & \sim_{t \rightarrow \infty} \ \ \exp\Big(-V_{\Lambda_\eta^\epsilon}(r) t\Big) .
\end{eqnarray}
$U(r;t_1,t_2)$ denotes a straight line of temporal gauge links at $r$ from time $t_1$ to $t_2$ and $\langle \ldots \rangle_U$ is the average on an ensemble of gauge link configurations distributed according to $e^{-S}$.

All computations presented in this work have been performed using SU(3) lattice gauge theory and the standard Wilson gauge action. We use a single ensemble of $5 \, 500$ gauge link configurations, generated with the Chroma QCD library \cite{Edwards:2004sx}. The lattice volume is $24^3 \times 48$. For the gauge coupling we choose $\beta = 6.0$, which corresponds to lattice spacing $a \approx 0.093 \, \textrm{fm}$, when identifying $r_0$ with $0.5 \, \textrm{fm}$. Standard smearing techniques are applied to the gauge links appearing in the correlation functions (\ref{EQN599}). Temporal gauge links are HYP2 smeared, spatial gauge links are APE smeared.


\section{\label{SEC445}Optimization of hybrid static potential creation operators and trial states}

Since the signal-to-noise ratio of correlation functions (\ref{EQN599}) decreases exponentially with respect to the temporal separation, it is essential to identify hybrid static potential creation operators, which generate trial states with large ground state overlap. This allows to extract hybrid static potentials at rather small temporal separations, where the signal-to-noise ratio is favorable.

\begin{figure}[htpb]
\begin{center}
\includegraphics[scale=0.85]{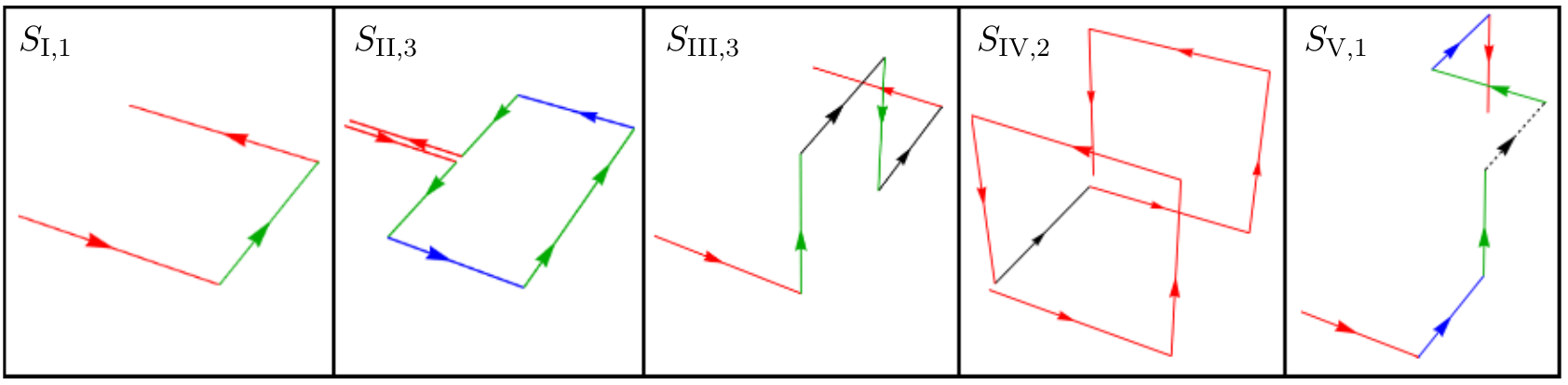}
\end{center}
\caption{\label{fig:0shapes}An exemplary subset of operators $S$ used to generate trial states $\ket{\Psi_\text{hybrid}}_{S;\Lambda_\eta^\epsilon}$ according to eq.\ (\ref{eq:trialstate}). Arrows represent straight paths of gauge links. Arrows with same color (red, green, blue) represent the same number of gauge links (black arrows can represent different numbers of gauge links). Dotted arrows can have length zero, while solid arrows represent at least one gauge link. 
}
\end{figure}

The starting point is a large set of 19 quite distinct operators $S$, where a small exemplary subset is shown in Figure~\ref{fig:0shapes}. From these operators we construct a large number of different trial states $\ket{\Psi_\text{hybrid}}_{S;\Lambda_\eta^\epsilon}$ using eq.\ (\ref{eq:trialstate}). To check, whether a trial state has large ground state overlap, we compute the effective mass
\begin{eqnarray}
\label{EQN632} V_{\textrm{eff};S;\Lambda_\eta^\epsilon}(r,t) a \ \ = \ \ \ln\bigg(\frac{W_{S,S;\Lambda_\eta^\epsilon}(r,t)}{W_{S,S;\Lambda_\eta^\epsilon}(r,t+a)}\bigg)
\end{eqnarray}
at temporal separation $t = a$, where contributions of excited states are most prominent. Small effective masses indicate trial states with large ground state overlaps, while operators leading to large effective masses can be discarded.

In a first step we consider each operator $S$ separately and optimize its extents for each hybrid potential sector $\Lambda_\eta^\epsilon$. In other words, for all $\Lambda_\eta^\epsilon$ and all arrows of all the example operators shown in Figure~\ref{fig:0shapes} we determine the number of gauge links they represent, such that the ground state overlap of the corresponding trial state is maximal.

To further improve the ground state overlaps of the trial states, we resort to variational techniques to generate the results shown in section~\ref{SEC499}. For each sector $\Lambda_\eta^\epsilon$ we select a small ``optimal set'' of 3 to 5 operators $S$, which yield the smallest effective masses at $t = a$ after the optimization outlined above. We then compute the corresponding correlation matrix and solve generalized eigenvalue problems.


\section{\label{SEC499}Numerical results}

We compute the ground state hybrid static potential for each of the sectors\\ $\Lambda_\eta^\epsilon = \Sigma_g^-,\Sigma_u^+,\Sigma_u^-,\Pi_g,\Pi_u,\Delta_g,\Delta_u$ as well as the ground state and first excited static potential for the sector $\Lambda_\eta^\epsilon = \Sigma_g^+$. For these computations we use correlation matrices $C_{j,k;\Lambda_\eta^\epsilon}(r,t) = W_{S_j,S_k;\Lambda_\eta^\epsilon}(r,t)$ and solve the generalized eigenvalue problems
\begin{eqnarray}
C_{\Lambda_\eta^\epsilon}(r,t) \mathbf{v}^{(n)}(r,t,t_0) \ \ = \ \ \lambda^{(n)}(r,t,t_0) C_{\Lambda_\eta^\epsilon}(r,t_0) \mathbf{v}^{(n)}(r,t,t_0)
\end{eqnarray}
with $t_0 = a$ and $n = 0,1,\ldots$. We then obtain the ground state potentials $V_{\Lambda_\eta^\epsilon}(r)$ by fitting a constant to the resulting effective potentials
\begin{eqnarray}
V_{\textrm{eff};\Lambda_\eta^\epsilon}^{(n)}(r,t,t_0) \ \ = \ \ \ln \frac{\lambda^{(n)}(r,t,t_0)}{\lambda^{(n)}(r,t+a,t_0)},
\end{eqnarray}
for $n = 0$ and similary the first excitation $V'_{\Sigma_g^+}(r)$ for $n = 1$.

\begin{figure}[htb]
\begin{center}
\includegraphics[scale=0.8]{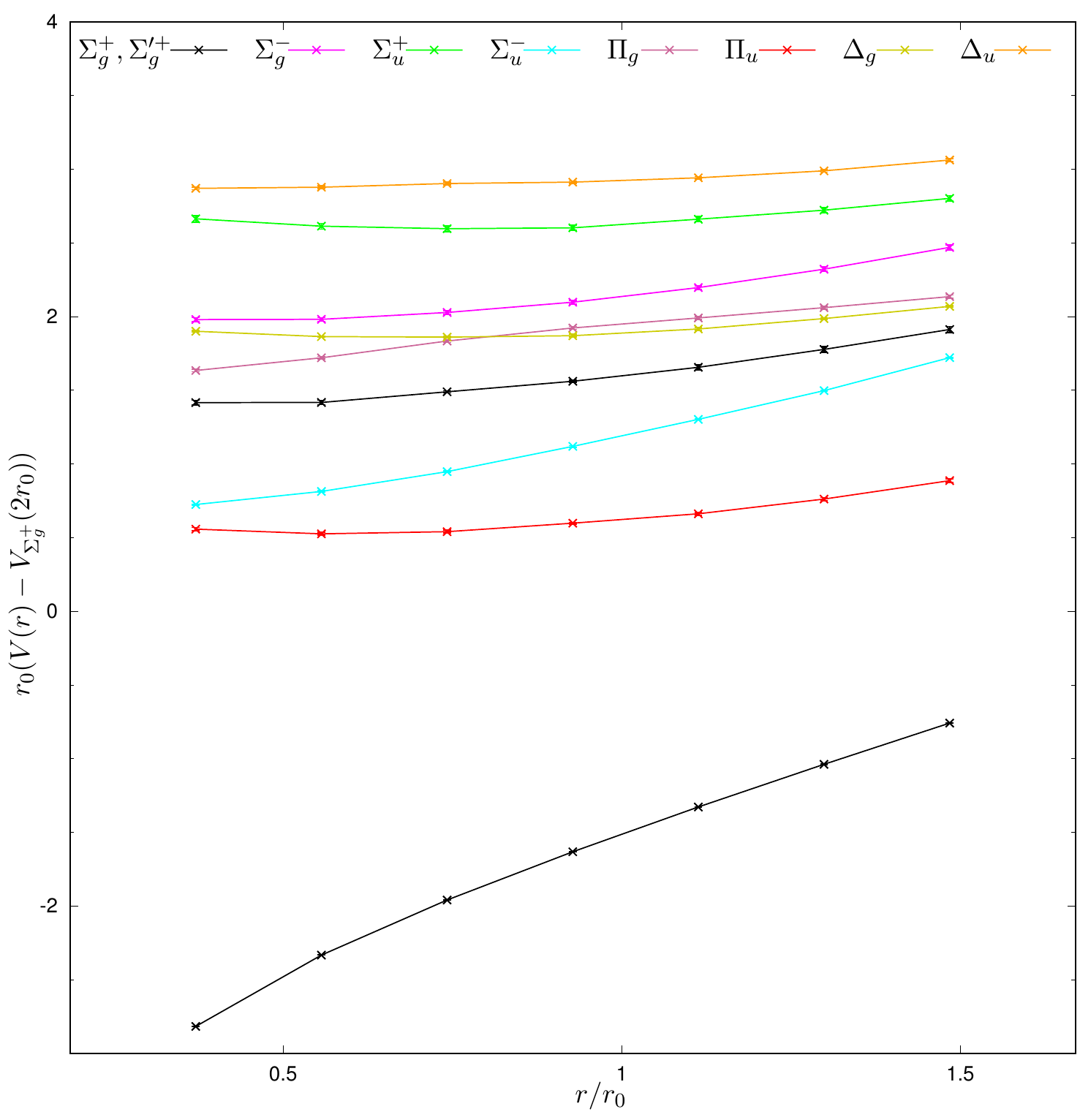}
\end{center}
\caption{\label{FIG002}The ordinary static potential $V_{\Sigma_g^+}(r) r_0$ and the corresponding first excitation $V'_{\Sigma_g^+}(r) r_0$ as well as the hybrid static potentials $V_{\Lambda_\eta^\epsilon}(r) r_0$, $\Lambda_\eta^\epsilon = \Sigma_g^-,\Sigma_u^+,\Sigma_u^-,\Pi_g,\Pi_u,\Delta_g,\Delta_u$ as functions of the separation $r / r_0$, where $r_0 = 0.5 \, \textrm{fm}$. To allow a straightforward comparison with results from the literature, e.g.\ with \cite{Juge:1997nc,Juge:2002br}, the vertical scale has been shifted by an additive constant such that $V_{\Sigma_g^+}(2 r_0) = 0$.}
\end{figure}

It is interesting to compare our resulting hybrid static potentials to the results from \cite{Juge:1997nc,Juge:2002br}, which are frequently used in recent publications (cf.\ e.g.\ \cite{Berwein:2015vca,Oncala:2017hop}) and seem to be the most accurate lattice results for hybrid static potentials currently available. For the majority of potentials there is no obvious qualitative discrepancy. Clearly visible differences can be observed for $V_{\Pi_g}(r)$ and $V_{\Delta_u}(r)$ at small separations $r$. Our results for these potentials are somewhat lower than those from \cite{Juge:2002br} and exhibit the expected approximate degeneracy with $V'_{\Sigma^+_g}(r)$ and $V_{\Sigma_u^+}(r)$, respectively (for a detailed discussion of these degeneracies and their relation to gluelump masses cf.\ e.g.\ \cite{Brambilla:1999xf}). Interestingly, we have found that the resulting potentials $V_{\Pi_g}(r)$ and $V_{\Delta_u}(r)$ are quite sensitive to the creation operators used in the correlation matrices. In both cases the operator $S_{IV,2}$ (cf.\ Figure~\ref{fig:0shapes}) significantly increases the ground state overlap and, thus, is essential to observe the previously mentioned and expected degeneracies at short $r$. We interpret this as indication that our selected sets of operators are better able to isolate the groundstate potentials for short $r$ in the $\Pi_g$ and $\Delta_u$ sectors than the operators used in \cite{Juge:1997nc,Juge:2002br}.


\section{Outlook}

An essential point of this work is the extensive optimization of hybrid static potential creation operators. We plan to use these optimized operators in future follow-up projects concerned with the computation of 3-point functions. Such 3-point functions might allow to compute quark spin corrections, to study decays to ordinary quarkonium states and glueballs or to investigate the gluon distribution inside heavy hybrid mesons. First results concerning the latter direction have recently been published \cite{Bicudo:2018yhk,Mueller:2018fkg,Bicudo:2018jbb}.


\section*{Acknowledgements}

C.R.\ acknowledges support by a Karin and Carlo Giersch Scholarship of the Giersch foundation. O.P.\ and M.W.\ acknowledge support by the DFG (German Research Foundation), grants PH 158/4-1 and WA 3000/2-1. M.W.\ acknowledges support by the Emmy Noether Programme of the DFG, grant WA 3000/1-1. This work was supported in part by the Helmholtz International Center for FAIR within the framework of the LOEWE program launched by the State of Hesse.

Calculations on the LOEWE-CSC and on the on the FUCHS-CSC high-performance computer of the Frankfurt University were conducted for this research. We would like to thank HPC-Hessen, funded by the State Ministry of Higher Education, Research and the Arts, for programming advice.



\end{document}